\documentclass[aps,twocolumn,superscriptaddress]{revtex4-2}

\usepackage{amsmath,amssymb,amsfonts,float,graphicx,tabularx,xcolor}

\usepackage[unicode=true,colorlinks=true]{hyperref}

\hypersetup{linkcolor=blue,citecolor=blue,urlcolor=blue}

\begin{document}

\title{Fractional Chern insulators in alternating twisted multilayer MoTe$_{2}$}


\author{Xi-Hang Feng}
\thanks{These two authors contributed equally.}

\author{Shi-Ping Ding}
\thanks{These two authors contributed equally.}

\author{Xiang-Jian Hou}

\author{Ying-Hai Wu}
\email{yinghaiwu88@hust.edu.cn}

\author{Jin-Hua Gao}
\email{jinhua@hust.edu.cn}
\affiliation{School of Physics and Wuhan National High Magnetic Field Center, Huazhong University of Science and Technology, Wuhan, Hubei 430074, China}
\affiliation{Hubei Fundamental Research Center for Physics, Wuhan, Hubei, China.}
 
\begin{abstract}
We study strongly correlated many-body states in alternating twisted trilayer and tetralayer MoTe$_{2}$. By sliding the top layer with respect to others and applying a perpendicular electric field, a variety of band structures can be realized. In many cases, the topmost hole band has unity Chern number and its quantum geometric properties can be tuned to some extent. Exact diagonalizations suggest that fractional Chern insulators are stabilized in certain parameter regimes but not in some regimes even when the band is topological. This contrast is attributed primarily to different quantum geometries as quantified by the trace condition. Our results demonstrate that sliding can serve as a useful knob for probing many-body states in moir\'e systems.
\end{abstract}
    
\maketitle
    
\section{Introduction}

Fractional quantum Hall (FQH) states have been a perpetual source of inspiration in the studies of strongly correlated many-body systems~\cite{Halperin-Book}. In the conventional setting, high quality two-dimensional electron systems are exposed to a perpendicular magnetic field. Landau quantization gives rise to extensive degeneracy in the single-particle orbitals such that many-body states are primarily determined by electron-electron correlations. From the topological perspective, Landau levels are not necessary for realizing FQH states. An energy band in two-dimensional lattice systems can be characterized by the Chern number~\cite{Thouless1982}. It is possible to have nonzero Chern number in the absence of external magnetic field~\cite{Haldane1988}. If the band is sufficiently flat, it would be a close analog of Landau level and can support fractional topological states in principle. These states are often referred to as fractional quantum anomalous Hall (FQAH) states or fractional Chern insulators (FCIs)~\cite{TangE2011,SunK2011,Neupert2011,ShengDN2011,Regnault2011,LiuZ2024}. 

For a large variety of carefully designed toy models, the existence of FCIs has been established by extensive numerical studies, but they do not seem to be relevant to solid state experiments. In recent year, experimental breakthroughs in van der Waals materials with moir\'e patterns reignite widespread interest in FCIs~\cite{Bernevig2025}. Experimental signatures of FCIs have been found in three types of systems: twisted bilayer MoTe$_{2}$~\cite{CaiJQ2023,ZengYH2023,ParkHJ2023,XuF2023,ParkHJ2026}, rhombohedral multilayer graphene with hBN substrate alignment~\cite{LuZG2024,XieJ2025,Aronson2025,Butler2026}, and twisted rhombohedral $(m+n)$-layer graphene~\cite{DongJW2025,LiZX2025}. Besides some states that closely resemble well-known FQH states in continuum Landau levels, phenomenon that do not have Landau level counterparts have also been reported. One question that has drawn considerable attention is how to understand and predict the robustness of FCIs for a given model. It has been proposed that some criteria regarding the quantum geometric tensor of energy bands can serve as good indicators for the appearance of FCIs~\cite{Parameswaran_2012_prb,Roy_2014_PRB,Dobardzic_2013_prb,Jackson_2015_NC,Bauer_2016_prb,Shavit_2024_prl,Cano_2026_arXiv}. In more general contexts, quantum geometry has also been investigated extensively~\cite{Torma2023,LiuTY2024,YuJB2025}.

In this work, we study alternating twisted multilayer moir\'e systems in which neighboring layers are rotated to opposite directions~\cite{AlBuhairan_2023_prb,Liang_2025_prb,Fedorko2025,Nakatsuji_2025_nature,Ding_2026_prb}. The physics of twisted bilayer MoTe$_{2}$ has been discussed in great details~\cite{Wu_2019_prl,Li_2021_prr,Reddy_2023_prb_FQAH,Reddy_2023_prb_phasediagram,Crepel_2023_prb,DongJK_2023_prl,Goldman_2023_prl,Qiu_2023_prx,Xiao_2024_prl,Morales_2024_prl,Lu_2024_prl,Cheng_2024_PNAS,Abouel2024,Jia_2024_prb,Yu_2024_prb,Sharma_2024_prb,Ahn_2024_prb,LiHQ2024,Reddy2024,XuC2025,Fujimoto2025,Wang_2025_prl,Chen_2025_nc,He_2025_arxiv,Chen_2026_science,LiBH2026}. It is natural to ask if more layers can lead to interesting new results. One additional knob in such cases is that the layers can be slide relative to each other. This has no effect in twisted bilayer systems for small twist angles but becomes crucial when there are more layers. The continuum models for trilayer and tetralayer systems are given in Sec.~\ref{model}. We slide the top layer relative to others and apply a perpendicular electric field. Numerical calculations are performed to unveil many-body phases in Sec.~\ref{result}. It is demonstrated that the band structure and quantum geometry can be tuned by sliding the top layer and applying a perpendicular electric field. This can be utilized to realize FCIs and competing phases. The paper is summarized and possible future topics are briefly discussed in Sec.~\ref{discuss}.

\begin{figure}[htbp]
\centering
\includegraphics[width=0.48\textwidth]{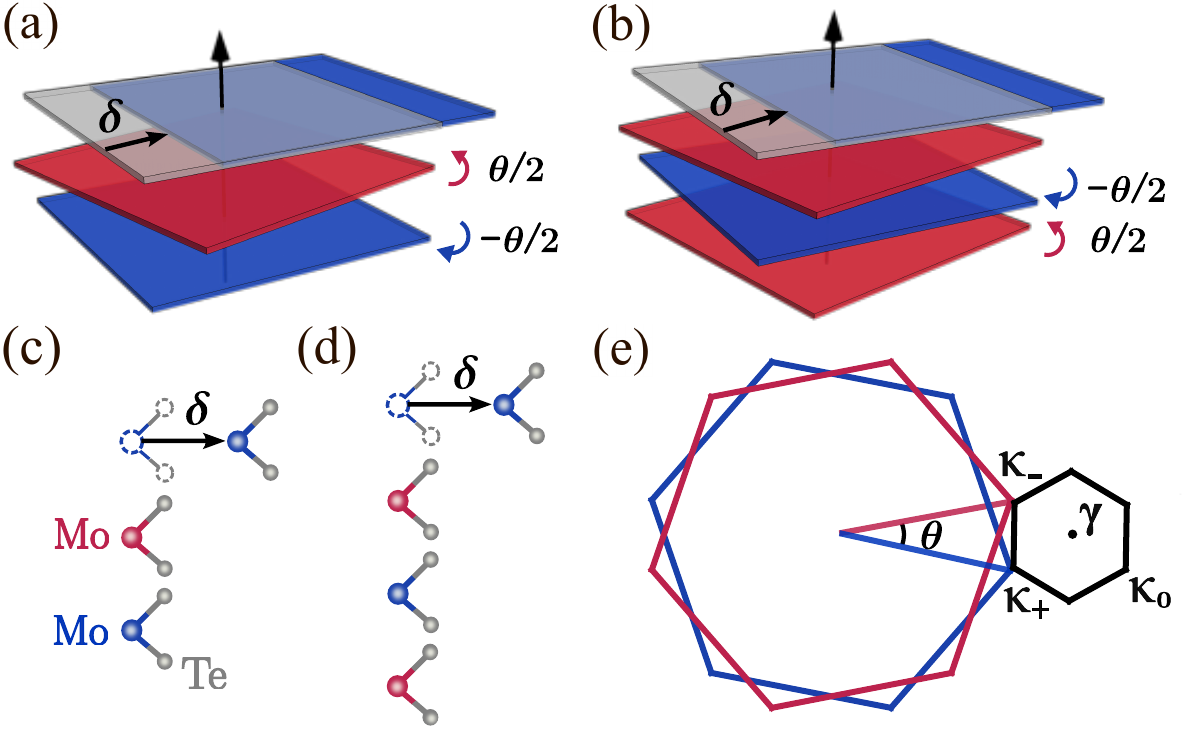}
\caption{(a,b) Schematics of alternating twisted trilayer and tetrlayer MoTe$_{2}$ in which the first layers are slid by $\boldsymbol{\delta}$. (c,d) Atomic side views of untwisted trilayer and tetralayer MoTe$_{2}$ with first layers slid. (e) Schematics of the Brillouin zones of individual layers and the moir\'e Brillouin zone.}
\label{fig:schematic}
\end{figure}

\section{Models and Methods}
\label{model}

We begin with the single-particle Hamiltonian for alternating twisted trilayer and tetralayer (AT3L and AT4L) MoTe$_{2}$. The system is shown schematically in Fig.~\ref{fig:schematic} (a) and (b). From top to bottom, the layers are labeled as $1,2,3,4$. The even layers are rotated in the counterclockwise direction by an angle $\theta/2$ whereas the odd layers are rotated in the opposite direction by an angle $\theta/2$ (indicated by a minus sign). In addition, the top layer is slid by a certain distance with respect to the other layers and a perpendicular electric field is applied to create different potential energies for the layers. For the cases without sliding, the band structure has been discussed by some of the authors~\cite{Liang_2025_prb}. In the absence of a displacement field, a symmetry decomposition was established: for an even (odd) number of layers, the system becomes equivalent to several copies of bilayer systems (and a monolayer with a moiré potential).

\begin{figure}[htbp]
\centering
\includegraphics[width=0.48\textwidth]{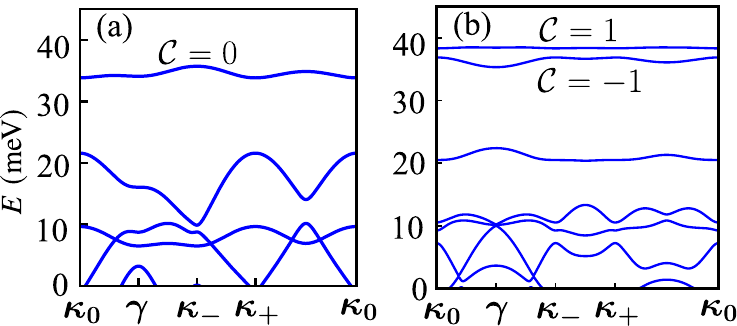}
\caption{Typical band structures of alternating twisted multilayer MoTe$_{2}$. (a) AT3L system with twist angle $2.15^{\circ}$; (b) AT4L system with twist angle $2.0^{\circ}$.}
\label{fig:bandex}
\end{figure}

In single layer transition metal dichalcogenides, spin and valley degrees of freedom are locked together, and each valley is described by a massive Dirac fermion~\cite{Xiao_2012_prl}. When multiple layers are stacked to generate moir\'e patterns, we follow the standard method to construct a low-energy continuum model~\cite{Bistritzer_2011_PNAS,Lopes_2012_prb}. The lattice constant of MoTe$_{2}$ is $a_{0} = 3.472 \, \mathring{\rm{A}}$ and the moir\'e lattice constant is $a_{M} \approx a_{0}/\theta$. The layers are labeled as $1,\cdots,L$. As shown in Fig.~\ref{fig:schematic} (e), the center of the moir\'e Brillouin zone is denoted as $\boldsymbol{\gamma}$, the momentum offset for the odd layers is 
\begin{equation}
\boldsymbol{\kappa}_{+} = \frac{4\pi}{3a_{M}} \left( -\frac{\sqrt{3}}{2} , -\frac{1}{2} \right),
\end{equation}
and that for the even layers is
\begin{equation}
\boldsymbol{\kappa}_{-} = \frac{4\pi}{3a_{M}} \left( -\frac{\sqrt{3}}{2} , \frac{1}{2} \right).
\end{equation}
The sliding vector of the $i$-th layer is denoted as $\boldsymbol{\delta}_{i}$, and the relative sliding vector between neighboring layers is $\boldsymbol{d}_{i}=(-1)^{i} (\boldsymbol{\delta}_{i+1}-\boldsymbol{\delta}_{i})$. There is a displacement field along the perpendicular direction so each layer acquires an electric potential energy.

In the $\boldsymbol{\kappa}_{+}$ valley, the single-particle Hamiltonian is
\begin{equation}
\mathcal{H}_{\rm AT} = \mathcal{H}_{\boldsymbol{k}}(\boldsymbol{\delta}_i) + \mathcal{V}_{\varepsilon},
\label{eq:SingleHam}
\end{equation}
where the crystal momentum $\boldsymbol{k}$ is measured relative to $\boldsymbol{\gamma}$. The second part of Eq.~\eqref{eq:SingleHam} is the layer-dependent electric potential
\begin{equation}
\mathcal{V}_{\varepsilon} = D \; \mathrm{diag}(0,1,\cdots,L-1),
\end{equation}
where the prefactor $D$ is determined by the displacement field, the interlayer distance, and the dielectric constant. The first part of Eq.~\eqref{eq:SingleHam} consists of kinetic energy, moir\'e potential, and interlayer tunneling
\begin{align}
\mathcal{H}_{\boldsymbol{k}}(\boldsymbol{\delta}_{i}) = 
\begin{bmatrix}
\mathcal{H}_{1}            & \mathcal{T}_{bm}            &  0      \\
\mathcal{T}_{bm}^{\dagger} & \widetilde{\mathcal{H}}_{m} &  \mathcal{T}_{tm} \\
0                & \mathcal{T}_{tm}^{\dagger}  &  \mathcal{H}_{L}
\end{bmatrix}.
\label{eq:PartI}
\end{align}
Its first and last diagonal components
\begin{equation}
\begin{aligned}
\mathcal{H}_{1} & = -\frac{\hbar^{2}}{2m^{*}} \left( \boldsymbol{k} - \boldsymbol{\kappa}_{1} \right)^{2} + \Delta_{1} \left(\boldsymbol{d}_{1}\right), \\
\mathcal{H}_{L} & = -\frac{\hbar^{2}}{2m^{*}} \left( \boldsymbol{k} - \boldsymbol{\kappa}_{L} \right)^{2} + \Delta_{L} \left(\boldsymbol{d}_{L-1}\right),
\end{aligned}
\end{equation}
are associated with the two outermost layers, where $\boldsymbol{\kappa}_{2n}=\boldsymbol{\kappa}_{-}$, $\boldsymbol{\kappa}_{2n-1}=\boldsymbol{\kappa}_{+}$ are the momentum offsets defined before and
\begin{equation}
\Delta_{i} (\boldsymbol{u}) = 2 V \sum_{j=1,3,5} \cos \left[ \boldsymbol{g}_{j} \cdot \boldsymbol{r} + \boldsymbol{G}_{j} \cdot\boldsymbol{u} + (-1)^{i+1} \psi \right]
\end{equation}
is the layer-dependent moir\'e potential. Here $\boldsymbol{g}_{1}=\left( 4\pi/\sqrt{3} a_{M} , 0 \right)$ is one moir\'e reciprocal lattice vector, $\boldsymbol{G}_{1}=(0, 4{\pi}a_{0}/\sqrt{3})$ is one first-shell reciprocal lattice vectors of single layer, and $\boldsymbol{g}_j$ ($\boldsymbol{G}_j$) is the vector obtained by $(j-1)\pi/3$ counterclockwise rotation of $\boldsymbol{g}_1$ ($\boldsymbol{G}_1$). We set the parameters as $m^{*} = 0.62 \, m_{e}$, $V=8$ meV, and $\psi=-89.6^{\circ}$~\cite{Wu_2019_prl,Jia_2024_prb,Reddy_2023_prb_FQAH,Mao2024}. 

To describe the other components in Eq.~\eqref{eq:PartI}, it is useful to define the kinetic-potential term
\begin{equation}
\mathcal{H}_{i} = -\frac{\hbar^{2}}{2m^{*}} \left( \boldsymbol{k}-\boldsymbol{\kappa}_{i} \right)^{2} + \Delta_{i} \left( \boldsymbol{d}_{i-1} \right) + \Delta_{i} \left(\boldsymbol{d}_{i}\right)
\end{equation}
and the tunneling term
\begin{eqnarray}
\mathcal{T}_{i,j} \left( \boldsymbol{\delta}_{i} , \boldsymbol{\delta}_{j} \right) = w \Big[ 1 + \mathrm{e}^{ -\mathrm{i} s \, \boldsymbol{g}_{2} \cdot \boldsymbol{r} } \mathrm{e}^{ -\mathrm{i} \boldsymbol{G}_{2} \cdot \left( \boldsymbol{\delta}_{i} - \boldsymbol{\delta}_{j} \right) } \notag\\
\qquad\qquad + \mathrm{e}^{-\mathrm{i} s \, \boldsymbol{g}_{3} \cdot \boldsymbol{r} } \mathrm{e}^{ - \mathrm{i} \boldsymbol{G}_{3} \cdot \left( \boldsymbol{\delta}_{i} - \boldsymbol{\delta}_{j} \right)} \Big],
\end{eqnarray}
where $w=-8.5$ meV, $i,j=2,\cdots,L-1$, $|i-j|=1$, and $s=(-1)^{i-1}$. It is obvious that $\mathcal{T}_{i,j} = \mathcal{T}_{j,i}^*$ when $|i-j|=1$. The diagonal component for the inner $L-2$ layers is
\begin{equation}
\widetilde{\mathcal{H}}_{m} = 
\begin{bmatrix}
\mathcal{H}_{2}     & \mathcal{T}_{2,3}   & 0        & \cdots &  0      \\
\mathcal{T}_{3,2}   & \mathcal{H}_{3}     & \mathcal{T}_{3,4}  & \cdots &  0      \\
0         & \mathcal{T}_{4,3}   & \mathcal{H}_{4}    & \cdots &  0 	 \\
\vdots    & \vdots    & \vdots   & \ddots &  \vdots   \\
0         & 0         & 0        & \cdots &  \mathcal{H}_{L-1}
\end{bmatrix}
\end{equation}
and the off-diagonal components are
\begin{equation}
\begin{aligned}
& \mathcal{T}_{bm} = \left( \mathcal{T}_{1,2}, 0, \ldots, 0 \right) \\
& \mathcal{T}_{tm} = \left( 0, \ldots, 0, \mathcal{T}_{L-1,L} \right)^{T}.
\end{aligned}
\end{equation}
We compute the band structure of Eq.~\eqref{eq:SingleHam} using standard method~\cite{Bistritzer_2011_PNAS}. Using the plane wave basis as $|\boldsymbol{k}+\boldsymbol{g},\ell\rangle$, the Bloch state of the moir\'e system can be written as
\begin{equation}
|\boldsymbol{k}\rangle = \sum_{\ell} \sum_{\boldsymbol{g}} z_{\boldsymbol{k},\boldsymbol{g},\ell} |\boldsymbol{k}+\boldsymbol{g},\ell\rangle,
\end{equation}
where $\boldsymbol{k}$ is the crystal momentum, $\boldsymbol{g}$ is a moir\'e reciprocal lattice vector within the cutoff shell, and the layers are denoted by $\ell\in[1,2,\cdots,L]$. Since the reciprocal lattice is infinite, we must introduce a truncation in practical calculations: the moir\'e reciprocal lattice vectors are restricted to $\boldsymbol{g}=n_1\boldsymbol{g}_1+n_2\boldsymbol{g}_2$ with $|n_i|\leq 6$. Two typical band structures are displayed in Fig.~\ref{fig:bandex}.

Next we turn to the many-body problem. For our present study, it is assumed that the electrons are spin-valley polarized so these indices shall be suppressed. The many-body Hamiltonian has two parts: the first one is a many-body kinetic energy term $H_{0}$ derived from the single-particle Hamiltonian Eq.~\eqref{eq:SingleHam} and the second one is an interaction term
\begin{equation}
H_{\mathrm{int}} = \frac{1}{2} \int \mathrm{d}\boldsymbol{r}_{1} \mathrm{d}\boldsymbol{r}_{2} \, : \rho(\boldsymbol{r}_{1}) V(\boldsymbol{r}_{1} - \boldsymbol{r}_{2}) \rho(\boldsymbol{r}_{2}) :.
\label{eq:ManyHam}
\end{equation}
Here $V(\boldsymbol{r}_{1} - \boldsymbol{r}_{2})$ is the interaction potential, $\psi^\dagger(\boldsymbol{r})$ is the creation operator at coordinate $\boldsymbol{r}$, $\rho(\boldsymbol{r})=\psi^{\dag}(\boldsymbol{r})\psi(\boldsymbol{r})$ is the density operator, and $:~:$ implements normal ordering. Numerical calculations are performed in finite size systems. The numbers of cells along the two primitive lattice directions are $N_{1}$ and $N_{2}$, respectively. The moir\'e unit cell has area $A_{M}$. The convention of Fourier transform is
\begin{eqnarray}
V(\boldsymbol{r}) = \frac{1}{A} \sum_{\boldsymbol{q}} V(\boldsymbol{q}) \exp\left( -\mathrm{i}\boldsymbol{q}\cdot\boldsymbol{r} \right)
\end{eqnarray}
with $A=N_{1}N_{2}A_{M}$ being the total area. 

As usually done in the literature, we project Eq.~\eqref{eq:ManyHam} to the topmost hole band. This leads to a second quantized Hamiltonian
\begin{align}
H & = \sum_{\boldsymbol{k}} \varepsilon_{\boldsymbol{k}} C^{\dag}_{\boldsymbol{k}} C_{\boldsymbol{k}} \notag \\
  &   \quad + \frac{1}{2} \sum_{\boldsymbol{k}'_{1},\boldsymbol{k}'_{2},\boldsymbol{k}_{1},\boldsymbol{k}_{2}} 
      V_{\boldsymbol{k}'_{1}\boldsymbol{k}'_{2};\boldsymbol{k}_{1}\boldsymbol{k}_{2}} C^{\dag}_{\boldsymbol{k}'_{1}}  C^{\dag}_{\boldsymbol{k}'_{2}} C_{\boldsymbol{k}_{2}} C_{\boldsymbol{k}_{1}}
\label{eq:ManyBodyHam}
\end{align}
that is expressed using the creation (annihilation) operator $C^{\dag}_{\boldsymbol{k}}$ ($C_{\boldsymbol{k}}$) for the momentum state $\boldsymbol{k}$. We define $\boldsymbol{q}_{\boldsymbol{g}} = \boldsymbol{k}'_{1}-\boldsymbol{k}_{1}+\boldsymbol{g}$ as the transfer of crystal momentum, $\boldsymbol{k}' \equiv [\boldsymbol{k}+\boldsymbol{q}]$ as the momentum obtained by folding $\boldsymbol{k}+\boldsymbol{q}$ back to the first moir\'e Brillouin zone, and $\boldsymbol{g}_{\boldsymbol{k},\boldsymbol{q}}$ as the reciprocal lattice vector satisfying $\boldsymbol{k}+\boldsymbol{q} = [\boldsymbol{k}+\boldsymbol{q}] + \boldsymbol{g}_{\boldsymbol{k},\boldsymbol{q}}$. The two-body matrix element is
\begin{equation}
V_{\boldsymbol{k}'_{1}\boldsymbol{k}'_{2};\boldsymbol{k}_{1}\boldsymbol{k}_{2}} = \frac{1}{A} \sum_{\boldsymbol{g}}
V(\boldsymbol{q}_{\boldsymbol{g}}) \Lambda(\boldsymbol{k}_{1},\boldsymbol{q}_{\boldsymbol{g}}) \Lambda(\boldsymbol{k}_{2},-\boldsymbol{q}_{\boldsymbol{g}})
\end{equation}
with form factor
\begin{equation}
\Lambda(\boldsymbol{k},\boldsymbol{q}) = \sum_{\ell} \sum_{\boldsymbol{g}} z^{*}_{[\boldsymbol{k}+\boldsymbol{q}],\boldsymbol{g}+\boldsymbol{g}_{\boldsymbol{k},\boldsymbol{q}},\ell} z_{\boldsymbol{k},\boldsymbol{g},\ell}.
\end{equation}
This quantity is nonzero only if the total crystal momentum is conserved up to a moir\'e reciprocal lattice vector
\begin{equation}
\boldsymbol{k}'_{1} + \boldsymbol{k}'_{2} = \boldsymbol{k}_{1} + \boldsymbol{k}_{2} \quad \mathrm{mod} \quad \boldsymbol{g}.
\end{equation}
In the Fourier transformed Coulomb potential $V(\boldsymbol{q})= 2{\pi}e^{2}/\left( \epsilon|\boldsymbol{q}| \right)$, the $\boldsymbol{q}=0$ component is canceled by an uniform positive charge background. 

\begin{figure}[H]
\centering
\includegraphics[width=0.45\textwidth]{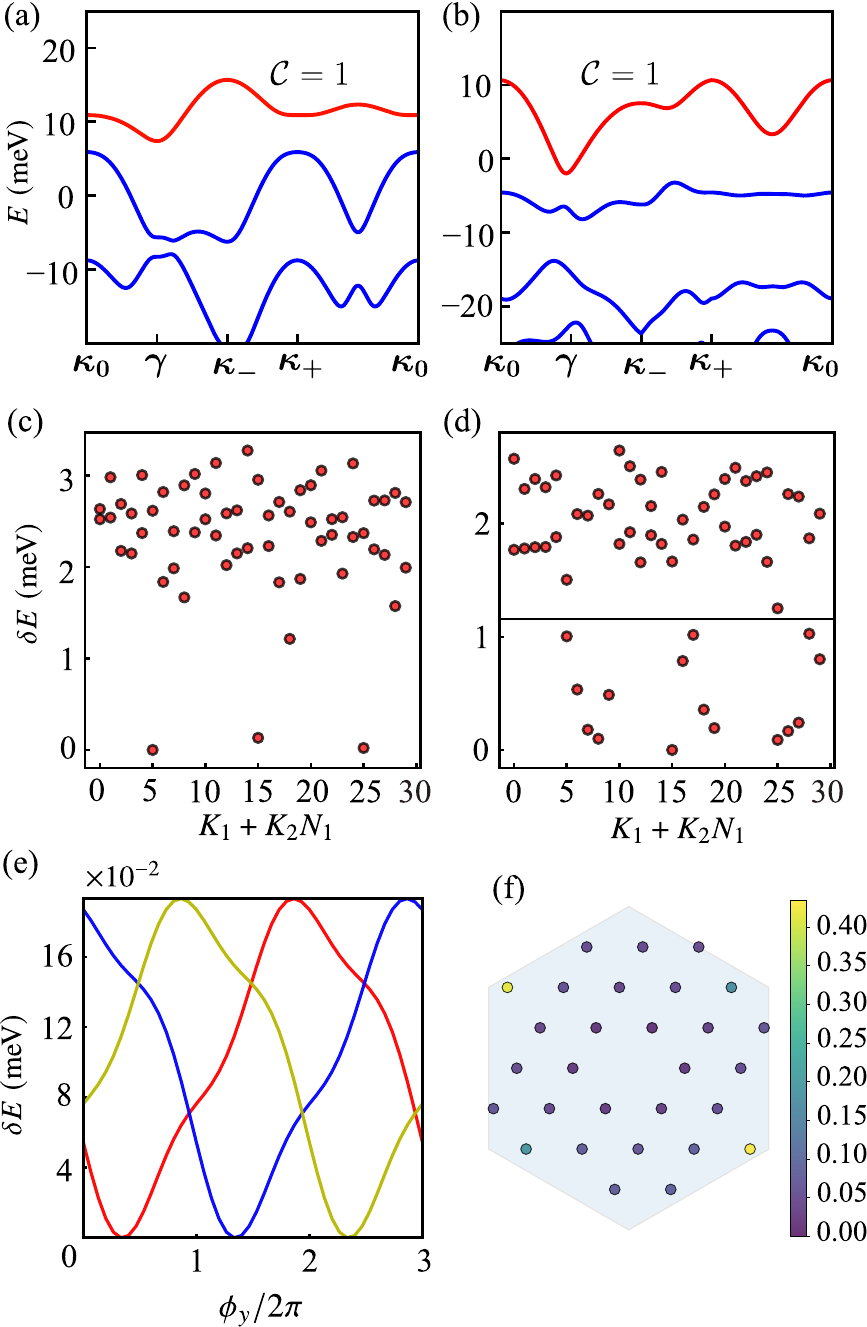}
\caption{(a,b) Band structures of AT3L MoTe$_{2}$ with $\theta=3.0^{\circ}$, $D=10$ meV, and (a) $\boldsymbol{\delta}=0$ (b) $\boldsymbol{\delta}=0.5\boldsymbol{a}_{1}$. (c) Many-body energy spectra at $1/3$ filling of the topmost band in (a). (d) The same plot as in (c) but for the topmost band in (b). The number of electrons is $10$ and the numbers of moi\'e cells are $5$ and $6$. (e) Spectral flow of the three quasi-degenerate ground states in (c) versus the boundary twist angle along the $y$ direction. (f) Static structure factor corresponding to (d).}
\label{FIG3}
\end{figure}

\begin{figure}[H]
\centering
\includegraphics[width=0.48\textwidth]{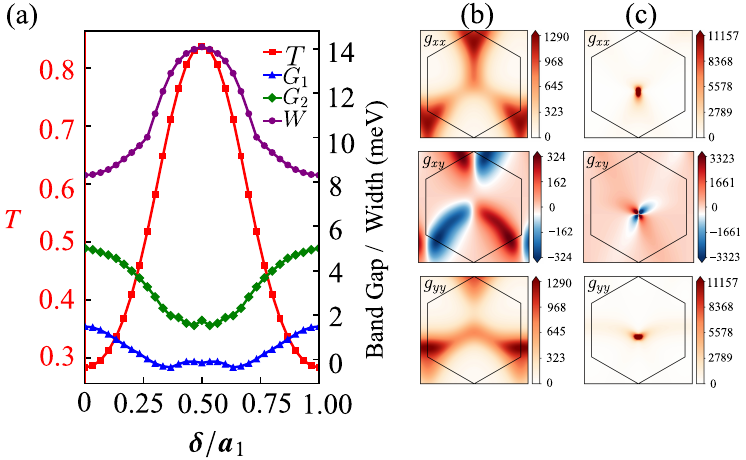}
\caption{(a) Evolution of the trace condition $T$, band width $W$, indirect band gap $G_{1}$, and direct band gap $G_2$ with the sliding distance $\boldsymbol{\delta}$ when the electric potential is fixed at $D=10$ meV. (b,c) Fubini-Study metric of AT3L MoTe$_{2}$ with $\theta=3.0^{\circ}$, $D=10$ meV, and (b) $\boldsymbol{\delta}=0$ (c) $\boldsymbol{\delta}=0.5\boldsymbol{a}_{1}$.}
\label{FIG4}
\end{figure}

\section{Numerical Results and Discussions}
\label{result}

We have studied AT3L and AT4L systems with various sliding distance $\boldsymbol{\delta}$ and electric potential energy $D$. The Chern number of an energy band is denoted as $\mathcal{C}$. In the trilayer system without sliding and electric field, the first hole band has $\mathcal{C}=0$ and no FCI is found. As the electric field gradually increases, the energy gap below the first hole band closes and reopens. For zero silding and $D=10$ meV, the first hole band has $\mathcal{C}=1$ as shown in Fig.~\ref{FIG3} (a). Exact diagonalization in this band unveiled good evidence of FCI. We denote the number of electrons as $N_{e}$. The many-body Hamiltonian is block diagonal with respect to the total crystal momenta along the two reciprocal lattice vectors. The eigenvalues are labeled using two integers $0 \leq K_{1} < N_{1}$ and $0 \leq K_{2} < N_{2}$ such that the total crystal momenta are $2{\pi}K_{1}/N_{1}$ and $2{\pi}K_{2}/N_{2}$. At filling factor $1/3$ ($N_{e}=10$, $N_{1}=5$, $N_{2}=6$), three quasi-degenerate ground states are observed in Fig.~\ref{FIG3} (c), whose total momenta agree with the generalized Pauli principle~\cite{Bernevig_2012_prb}. When boundary twist angle along one direction is introduced, spectral flow among these states are induced as shown in Fig.~\ref{FIG3} (e). These results suggest that a FCI is realized. 

If the top layer is slid by $0.5\boldsymbol{a}_{1}$, we obtain the band structure in Fig.~\ref{FIG3} (b). The first band still has $\mathcal{C}=1$ but single band approximation is less valid in this case. Exact diagonalization is useful but the results should be interpreted with care. At filling factor $1/3$, the system likely forms a charge density wave (CDW) instead of a FCI. Below the horizontal line in Fig.~\ref{FIG3} (d), there are 15 states at total momenta consistent with a CDW interpretation~\cite{Scarola_2015_prb,peng2026}. Ideally, these states should be degenerate but there are considerable splitting here. We have also calculated the static structure factor $\boldsymbol{S}(\boldsymbol{q})$~\cite{Dobardzic_2013_prb,Fu_2025_prb,Sharma_2024_prb}. For a robust CDW, it should exhibit high peaks at the ordering wave vector. In our case, there are two shallow peaks at opposite momenta. The ratio between the largest and second largest values is not very large, but still consistent with a CDW interpretation. 

The sharp difference between these two cases one can be attributed to the evolution of band width and quantum geometry with sliding. As one can see in Fig.~\ref{FIG4} (a), the band width increase with $\delta$ and reaches a peak at $0.5\boldsymbol{a}_{1}$, which is expected to suppress FCI. To characterize quantum geometric properties, we compute the tensor 
\begin{eqnarray}
Q_{\mu\nu} &=& \langle \partial_{\mu} \boldsymbol{k} | \partial_{\nu} \boldsymbol{k} \rangle - \langle \partial_{\mu} \boldsymbol{k} | \boldsymbol{k} \rangle \langle \boldsymbol{k} | \partial_{\nu} \boldsymbol{k} \rangle \nonumber \\
&=& g_{\mu\nu}(\boldsymbol{k}) - \frac{i}{2} \Omega_{\mu\nu}(\boldsymbol{k}),
\end{eqnarray}
where the real part $g_{\mu\nu}(\boldsymbol{k})$ is the quantum Fubini-Study metric [see Fig.~\ref{FIG4} (b,c)] and the imaginary part $\Omega_{\mu\nu}(\boldsymbol{k})$ is the Berry curvature. It has been proposed that the stability of FCI is related to the quantity~\cite{Parameswaran_2012_prb,Roy_2014_PRB,Parker2026}
\begin{equation}
T = \frac{1}{2\pi} \int \mathrm{d}\boldsymbol{k} \, \left\{ {\rm Tr} \Big[ g(\boldsymbol{k}) \Big] - \left| \Omega_{xy}(\boldsymbol{k})\right| \right\} .
\end{equation}
The lowest Landau level and some ideal flat bands have $T=0$ and they host robust fractional states. In contrast, large values of $T$ tend to destroy FCIs. This is referred to as the trace condition. As shown in Fig.~\ref{FIG4} (a), $T$ increases considerably when the sliding distance changes from $0$ to $0.5\boldsymbol{a}_{1}$, which should also play a role in destruction of FCI. It is unfortunate that the band width and $T$ increase concomitantly, so their effects cannot be separated. To this end, we have tested a toy model in which the topmost hole band is artificially flattened (i.e., the single-particle energy at all momenta are neglected). While the band width is reduced to zero, the single-particle eigenstates are still the same, so quantum geometric properties are not affected. We still obtain a FCI at $\boldsymbol{\delta}=0$ and a CDW at $\boldsymbol{\delta} = 0.5\boldsymbol{a}_{1}$ (see the Appendix for details). This finding provides strong support for the claim that the trace condition is a good indicator of FCI.

\begin{figure}[h]
\centering
\includegraphics[width=0.45\textwidth]{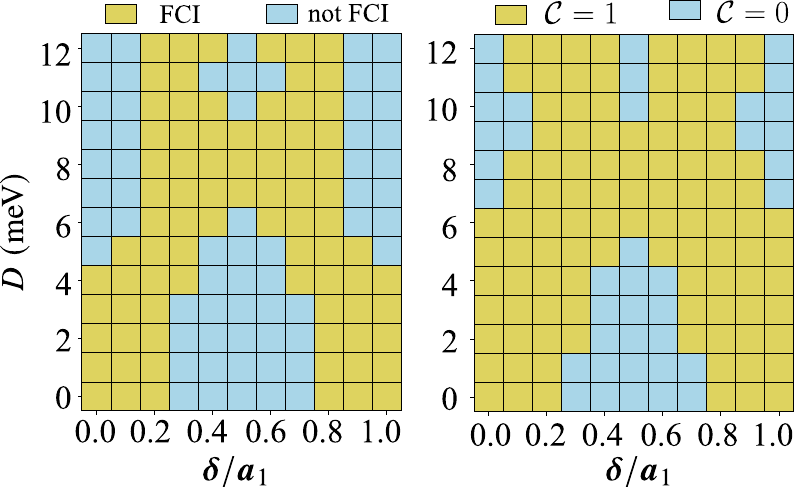}
\caption{Phase diagram of AT4L MoTe$_{2}$ with $\theta=2.5^{\circ}$ at filling factor $1/3$. Numerical calculations were performed at many different combinations of electric potential $D$ and sliding distance $\boldsymbol{\delta}$. The number of electrons is $8$ and the numbers of moir\'e cells are $4$ and $6$. FCIs are identified based on ground state degeneracy and spectral flow under flux insertion.}
\label{FIG5}
\end{figure}

Next we turn to tetralayer systems. In the absence of sliding and the displacement field, there is an isolated band with unity Chern number for twist angle $\theta \in [2.0^{\circ}, 4.0^{\circ}]$. For twist angles $\theta=2.5^{\circ}$ and $3.0^{\circ}$, numerical calculations at $\nu=1/3$ of the topmost hole band reveal clear FCI signatures (see the Appendix for details). This is broadly in line with expectation given that a AT4L system can be decomposed to two copies of twisted bilayer~\cite{Liang_2025_prb}. Furthermore, we have studied many combinations of nonzero $\boldsymbol{\delta}$ and $D$ at $\theta=2.5^{\circ}$. The phase diagram at $1/3$ filling ($N_{e}=8, N_{1}=4, N_{2}=6$) is displayed in Fig.~\ref{FIG5}. While the lower left corner hosts a FCI, it is destroyed by an intermediate $\boldsymbol{\delta}$ or a sufficiently large $D$. In some cases, the topmost hole band has $\mathcal{C}=0$ so the absence of FCI is natural. The fact that a topological band may not support a FCI is more interesting.

\begin{figure}[h]
\centering
\includegraphics[width=0.45\textwidth]{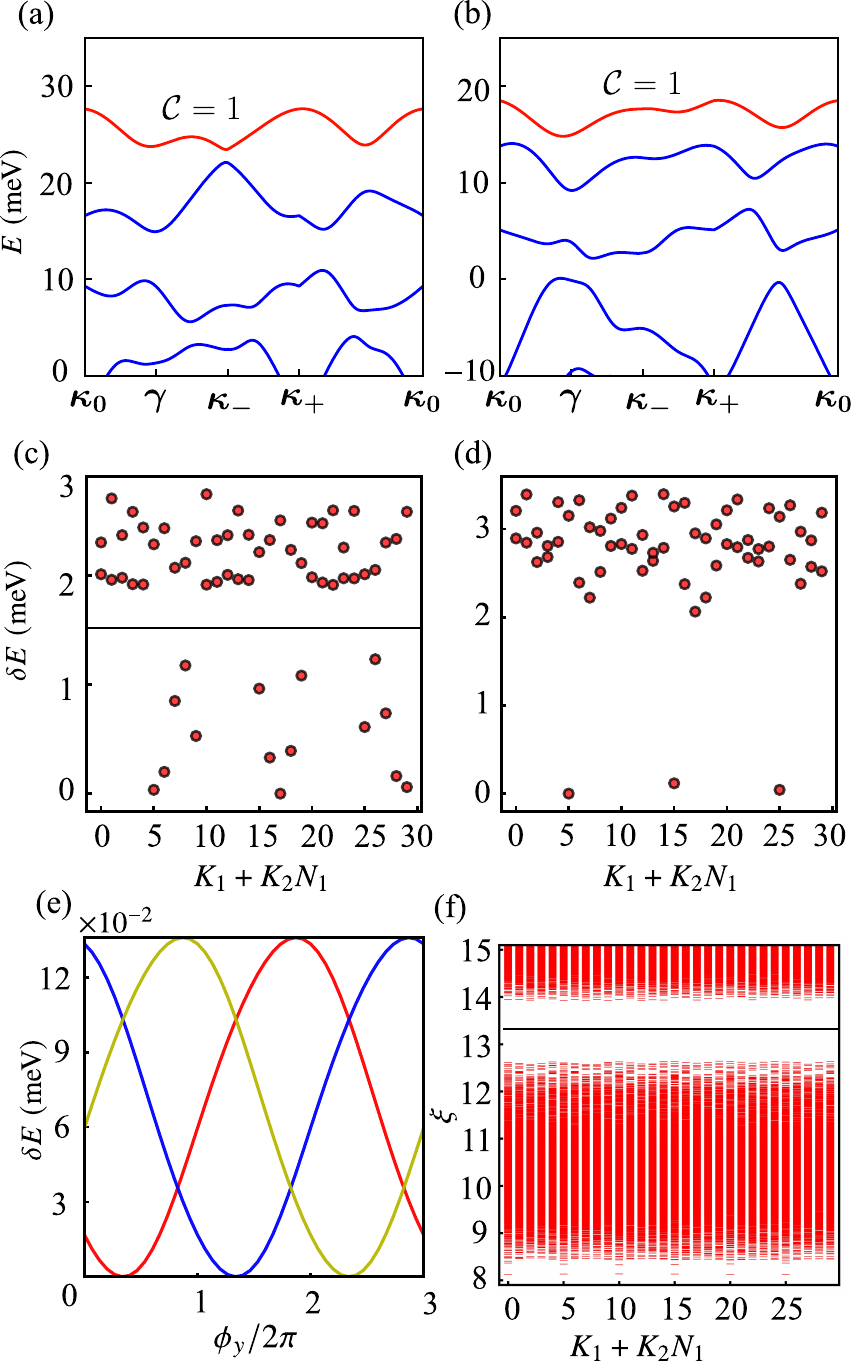}
\caption{(a,b) Band structures of AT4L MoTe$_{2}$ at $\theta=2.5^{\circ}$, $\boldsymbol{\delta}=0.3\boldsymbol{a}_{1}$, and (a) $D=2$ meV (b) $D=7$ meV. (c,d) Many-body energy spectra at $1/3$ filling of the topmost band in (a,b). The number of electrons is $10$ and the numbers of moir\'e cells are $5$ and $6$. (e) Spectral flow of the three quasi-degenerate ground states in (d) versus the boundary twist angle along the $y$ direction. (f) Particle entanglement spectrum of the three quasi-degenerate ground states in (d).}
\label{FIG6}
\end{figure}

An inspection of line cuts in the phase diagram unveils quantum geometric effects. Specifically, we focus on the cases with $\boldsymbol{\delta}=0.3\boldsymbol{a}_{1}$ and varying $D$. The single-particle band structures at $D=2$ meV and $7$ meV are displayed in Fig.~\ref{FIG6} (a,b). We project the many-body Hamiltonian to the topmost band and obtain the energy spectra at $1/3$ filling in Fig.~\ref{FIG6} (c,d). The case with $D=7$ meV realizes a FCI whereas the case with $D=2$ meV does not, as manifested by the ground state quasi-degeneracies. As shown in Fig.~\ref{FIG6} (e), when twist boundary condition is applied along one direction, the three quasi-degenerate ground states flow to each other and return to themselves after $6\pi$ fluxes are inserted. Using the three quasi-degenerate ground states, the particle entanglement spectrum is computed~\cite{Sterdyniak2011}. There is a clear entanglement gap in Fig.~\ref{FIG6} (f), and the counting of levels below the horizontal line agrees with theoretical prediction based on generalized Pauli principle and Brillouin zone folding~\cite{Bernevig_2012_prb}. By analyzing the quantum numbers of the states below the horizontal line in Fig.~\ref{FIG6} (c), we believe that the system realizes a CDW state at $D=2$ meV. As in the trilayer case, both band width and quantum geometry varies with $D$. Further calculations using the flat band Hamiltonian suggest that the destruction of FCI is primarily due to the sharp peak of $T$ (see the Appendix).

\begin{figure}[h]
\centering
\includegraphics[width=0.48\textwidth]{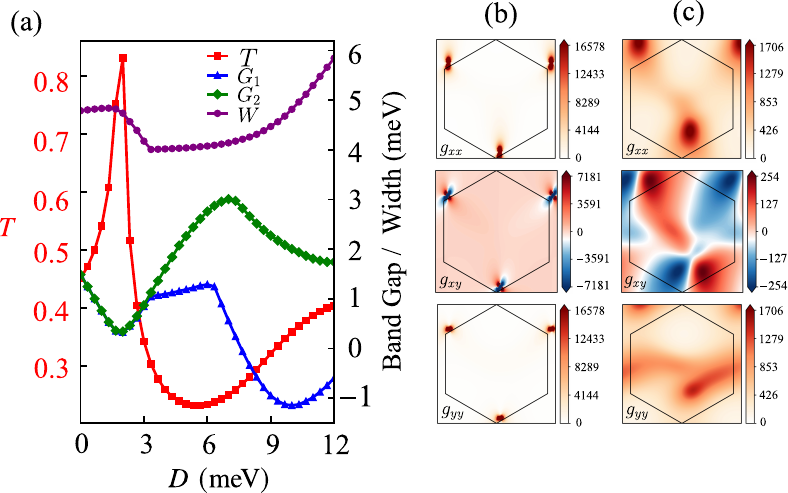}
\caption{(a) Evolution of the trace condition $T$, band width $W$, indirect band gap $G_{1}$, and direct band gap $G_{2}$ with the electric potential $D$ when the sliding distance is fixed at $\boldsymbol{\delta}=0.3\boldsymbol{a}_{1}$. (b,c) Fubini-Study metric of AT4L MoTe$_{2}$ with $\theta=2.5^{\circ}$, $\boldsymbol{\delta}=0.3\boldsymbol{a}_1$, and (b) $D=2$ meV (c) $D=7$ meV.}
\label{FIG7}
\end{figure}

\section{Summary}
\label{discuss}

In summary, we have demonstrated that sliding could serve as a useful tuning knob for studying strongly correlated states in moir\'e flat bands. When sliding is combined with displacement field, the properties of alternating twisted multilayer systems can be tuned to study many-body phases. This distinguishes systems with more than two layers with the bilayer ones. It is particularly intriguing that sliding provides a realistic method to change quantum geometry, thereby paves the way for a quantitative experimental test of theoretical predictions regarding its fundamental role in many-body physics.

There are plenty other questions that could be investigated. Besides the $1/3$ state that we focused on, Jain FCI states and gapless CFL states may also emerge under suitable conditions~\cite{Jain1989,Halperin1993}. On top of the ground states, elementary excitations such as anyons, rotons, and fractional excitons can be created~\cite{LiuZ2025,Kousa2025,Paul2025}. In higher hole bands, special combinations of form factor and quantum geometry can generate non-Abelian states~\cite{Ahn_2024_prb,Reddy2024,Wang_2025_prl,Chen_2025_nc,LiBH2026}. These topics should be examined in alternating twisted multilayer systems with sliding. While this work studied the cases of sliding the first layer along one direction, there is a vast parameter space corresponding to multiple layers being slid along different directions. We hope to report other interesting results in the future. 

\vspace{1em}

{\bf Note added} --- During the preparation of this manuscript, we became aware of two preprints about alternating twisted moir\'e systems~\cite{ZhangXW2026,Beach2026}.

\section*{Acknowledgments}

XHF thanks Aidan Reddy for helpful correspondence. SPD and JHG was supported by the National Key Research and Development Program of China (Grants No.~2022YFA1403501) and National Natural Science Foundation of China under Grant Nos.~12474169 and 12141401.

\newpage
\onecolumngrid

\appendix 

\section{Additional Results}

\begin{figure}[H]
\centering
\includegraphics[width=0.95\textwidth]{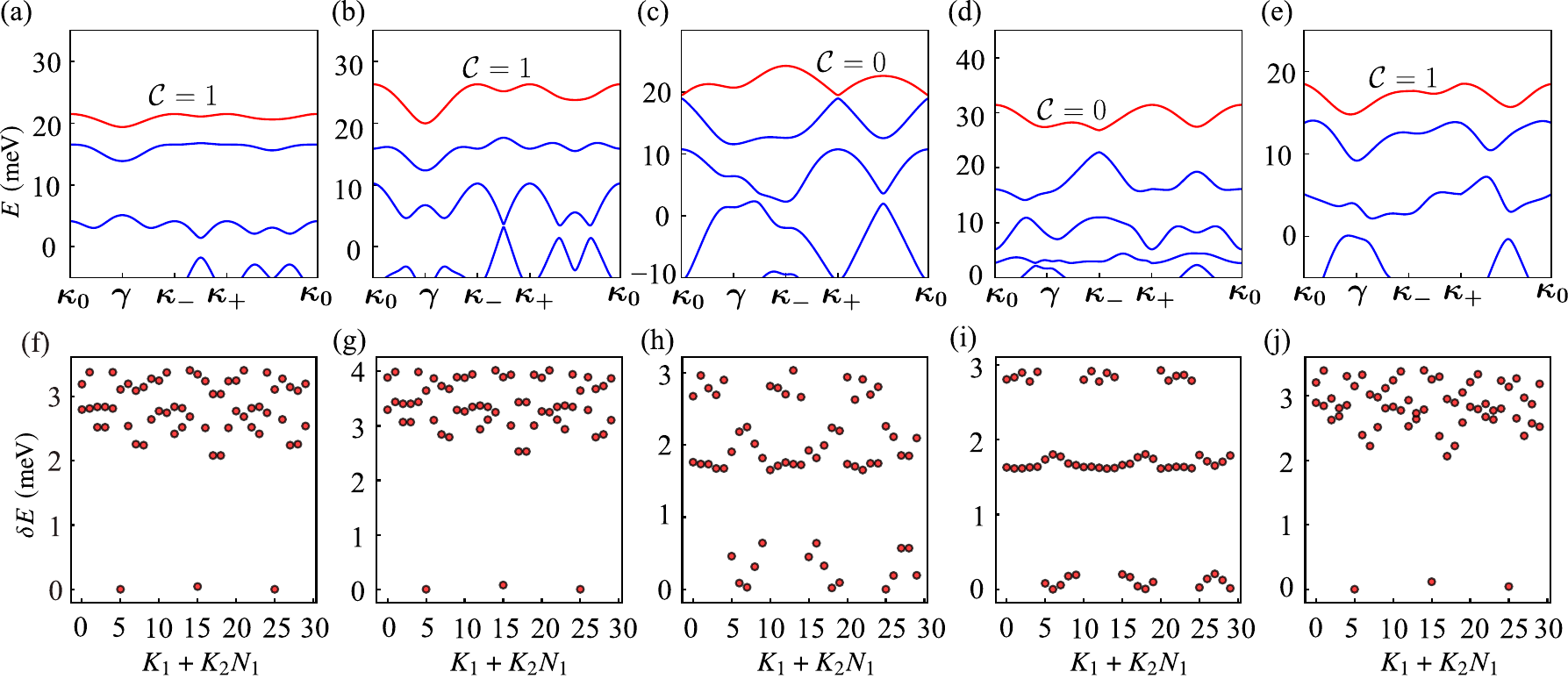}
\caption{The first row is band structures for AT4L MoTe$_{2}$ and the second row is the many-body energy spectra at $1/3$ filling of the topmost band. (a,f) $\theta=2.5^{\circ}$, $\boldsymbol{\delta}=0$, and $D=0$ meV. (b,g) $\theta=3.0^{\circ}$, $\boldsymbol{\delta}=0$, and $D=0$ meV. (c,h) $\theta=2.5^{\circ}$, $\boldsymbol{\delta}=0$, and $D=7$ meV. (d,i) $\theta=2.5^{\circ}$, $\boldsymbol{\delta}=0.5\boldsymbol{a}_{1}$, and $D=0$ meV. (e,j) $\theta=2.5^{\circ}$, $\boldsymbol{\delta}=0.3\boldsymbol{a}_{1}$, and $D=7$ meV. The number of electrons is $10$ and the numbers of moi\'e cells are $5$ and $6$.}
\label{FIGA1}
\end{figure}

\begin{figure}[H]
\centering
\includegraphics[width=0.5\linewidth]{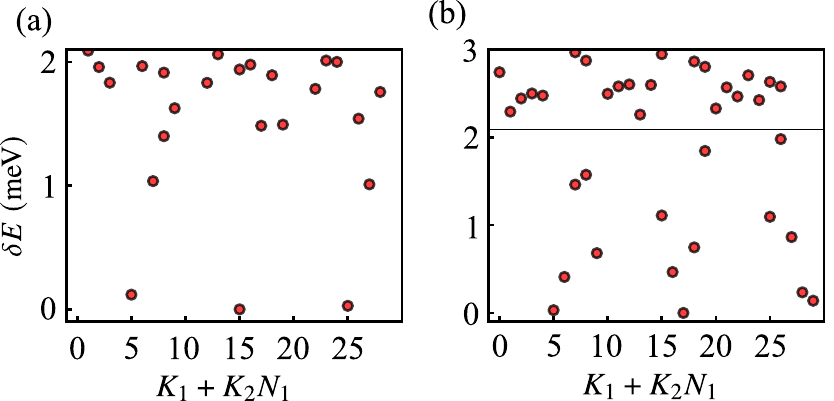}
\caption{Many-body energy spectra of AT3L with $\theta=3.0^{\circ}$, $D=10$ meV, and (a) $\boldsymbol{\delta}=0$ (b) $\boldsymbol{\delta}=0.5\boldsymbol{a}_{1}$. The flat band Hamiltonian is used and the system parameters are $N_{e}=10, N_{1}=5, N_{2}=6$.}
\label{FIGA2}
\end{figure}

\begin{figure}[H]
\centering
\includegraphics[width=0.5\linewidth]{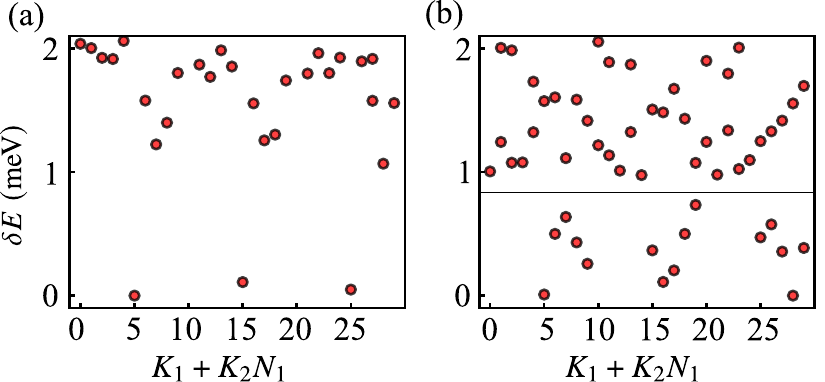}
\caption{Many-body energy spectra of AT4L with $\theta=2.5^{\circ}$, $\boldsymbol{\delta}=0.3\boldsymbol{a}_{1}$, and (a) $D=2$ meV (b) $D=7$ meV. The flat band Hamiltonian is used and the system parameters are $N_{e}=10, N_{1}=5, N_{2}=6$.}
\label{FIGA3}
\end{figure}

In this Appendix, we provide additional results. For AT4L MoTe$_{2}$, we have calculated the many-body energy spectra for a variety of cases. Fig.~\ref{FIGA1} presents some results for twist angles $\theta=2.5^{\circ}$ and $3.0^{\circ}$. When the topmost band is topologically trivial, the existence of CDW states can be seen more clearly. For AT3L MoTe$_{2}$ with $\theta=3.0^{\circ}$ and $D=10$ meV, we have shown that sliding causes destruction of FCI. To further separate the effects of band width and quantum geometry, we consider a modified flat band Hamiltonian, which drops the first term in Eq.~\eqref{eq:ManyBodyHam} and keeps only the second term. As one can see from Fig.~\ref{FIGA2}, the two cases are still FCI and non-FCI. It is thus reasonable to attribute their difference to quantum geometric effects. For AT4L MoTe$_{2}$ with $\theta=2.5^{\circ}$ and $\boldsymbol{\delta}=0.3\boldsymbol{a}_{1}$, flat band Hamiltonian has also been studied. The results presented in Fig.~\ref{FIGA3} also demonstrate the importance of trace condition. 

\bibliography{citation}

\end{document}